\documentclass[nofootinbib,article,
 reprint,
 amsmath,amssymb,
 aps,
]{revtex4-1}

\usepackage{filecontents}
\usepackage{graphicx}
\usepackage{dcolumn}
\usepackage{bm}
\usepackage{subfigure}
\usepackage{graphicx}

\begin{document}

\preprint{APS/123-QED}

\title{A Two-band Model for p-wave Superconductivity}

\author{Heron Caldas}
\author{F. S. Batista}
\affiliation{Departamento de Ci\^{e}ncias Naturais, Universidade Federal de S\~{a}o Jo\~{a}o Del Rei, Pra\c{c}a Dom Helv\'{e}cio 74, 36301-160, S\~{a}o Jo\~{a}o Del Rei, MG, Brazil\\}

\author{Mucio A. Continentino}
\affiliation{Centro Brasileiro de Pesquisas F\'{i}sicas, Rua Dr. Xavier Sigaud 150, 22290-180, Rio de Janeiro, RJ, Brazil}

\author{Fernanda Deus}
\affiliation{Universidade do Estado do Rio de Janeiro, Faculdade de Tecnologia, Departamento de Matem\'{a}tica, F\'{i}sica e Computa\c{c}\~{a}o, Rodovia Presidente Dutra km 298, 27537-000, Resende, RJ, Brazil}

\author{David Nozadze}
\affiliation{Cisco Systems, Inc., San Jose, CA, 95134, USA}

\date{\today}

\begin{abstract}
In this paper we study the effects of hybridization in the superconducting properties of a two-band system. We consider the cases that these bands are formed by electronic orbitals with angular momentum, such that, the hybridization $V(\mathbf{k})$ among them can be symmetric or antisymmetric under inversion symmetry.  We take into account only intra-band attractive interactions in the two bands and investigate the appearance of an induced inter-band pairing gap. We show that (inter-band) superconducting orderings are induced in the total absence of attractive interaction between the two bands, which turns out to be completely dependent on the hybridization between them. For the case of antisymmetric hybridization we show that the induced inter-band superconductivity has a p-wave symmetry.

\end{abstract}

\pacs{Valid PACS appear here}
\maketitle


\section{\label{intro}Introduction}

Some materials such as magnesium diboride ($\rm{MgB_{2 }}$) have unusual superconducting properties as, for example, high transition temperature $T_c$ and anomalous specific heat. Recent experimental investigations report the existence of superconductivity at $39~ \rm{K}$ in $\rm{MgB_{2 }}$~\cite{exp1}. In addition, other experimental studies have indicated that $\rm{MgB_{2 }}$ has two distinct superconducting gaps~\cite{exp2,exp3,exp4,exp5,exp6,exp7,exp8}. It has been shown in Ref.~\cite{Choi}, that the $\rm{MgB_{2 }}$ Fermi surface (FS) is determined by three orbitals, however only two different BCS gaps are experimentally observed. This happens because two of the three orbitals hybridize with each other and determine one single band, responsible for a large superconducting gap on the $\sigma$ FS, while the non-hybridized orbital determines a smaller superconducting gap at the $\pi$ band FS.

Consequently, the number of different gaps that arise in multi-orbital systems is directly related to the level of hybridization among the orbitals present in a given material~\cite{Moreo}. In this way it turns out to be clear that hybridization plays an important role in the physics of multi-band superconductors~\cite{hybri1,hybri2,hybri3,hybri4,hybri5,hybri6,hybri7,hybri8}. 

Notice that the hybridization can be symmetric or antisymmetric. It has been shown that symmetric ($k$-independent) hybridization acts in detriment of intra-band superconductivity~\cite{sym1,sym2}. Surprisingly, it has shown recently that antisymmetric ($k$-dependent) hybridization enhances superconductivity~\cite{anti1}.

\begin{figure}
\centering
 \includegraphics[width=8cm]{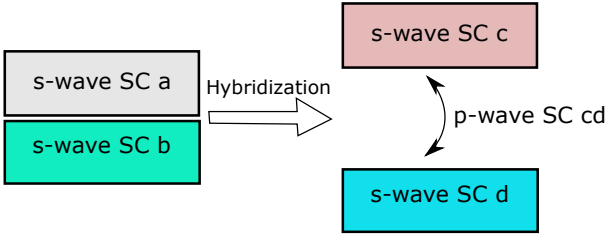}
\caption{(Color online) Setup proposed for realizing an induced p-wave pairing gap with the hybridization of two bands, say $a$ and $b$. Inside each of these bands there is intra-band ($s$-wave) superconductivity (SC) with pairing gaps $\Delta_a$ and $\Delta_b$. After the application of an antisymmetric hybridization $V(k)$, which may be pressure or doping, these intra-bands $s$-wave SC still exist, but now they are named $c$ and $d$ since they depend on $\Delta_a$, $\Delta_b$, and $V(k)$. Besides, there is the emergence of an induced $p$-wave SC with pairing gap $\Delta_{cd}$.}
\label{fig0}
\end{figure}

In this paper we study the emergence of an (inter-band) induced $p$-wave gap due to the hybridization of two single bands, say $a$ and $b$. We consider superconducting interactions only inside each band, which will result in intra-band pairing gaps $\Delta_a$ and $\Delta_b$, respectively, in these bands. We take into account symmetric and antisymmetric, k-dependent hybridization $V(k)$, and show that the latter produces an odd-parity mixing between the $a$ and $b$ bands, and will be responsible for the $p$-wave nature of the induced inter-band gap. See Fig.~\ref{fig0}.

Our main interest here is the controlled generation (from conventional $s$-wave superconductors) of a $p$-wave pairing gap, motivated by its recent remarkable features. $p$-wave one-dimensional superconductors have attracted considerable attention since its ``discovery'' by Kitaev~\cite{Kitaev1,Kitaev2}, which demonstrated that this system can host Majorana fermions (or, more precisely, zero energy bound states) at this ends. Due to the fact that the (condensed matter) Majorana fermions are topologically protected and they satisfy a criterion of robustness for use in topological quantum computers, they are promising candidates to act as q-bits~\cite{Ladd}.

\section{\label{model} Model Hamiltonian}

Here we introduce the system describing the hybridization of $s$ and $p$-bands. This system can be realized in different scenarios, as in condensed matter or in the cold atom context, where a bipartite lattice with $s$ and $p$-orbitals in different sub-lattices hybridize~\cite{Mart1,Mart2}.

We write the Hamiltonian of a 3D {\it effective} and generic two-band superconductor model in second quantization

\begin{align}
\label{eq:Hab}
H_{ab}  & =  \sum_{{\bf k}\sigma} \varepsilon_a ({\bf k}) a^\dagger_{{\bf k}\sigma} a_{{\bf k}\sigma} + \sum_{\bf k} \Delta_a({\bf k}) a^\dagger_{{\bf k}\uparrow} a^\dagger_{-{\bf k}\downarrow} +{\rm h.c.} \nonumber  \\
& +\sum_{{\bf k}\sigma} \varepsilon_b ({\bf k}) b^\dagger_{{\bf k}\sigma} b_{{\bf k}\sigma} +  \sum_{\bf k} \Delta_b({\bf k}) b^\dagger_{{\bf k}\uparrow} b^\dagger_{-{\bf k}\downarrow} +{\rm h.c.}  \nonumber \\ 
& +  \sum_{{\bf k}\sigma} V({\bf k}) a^\dagger_{{\bf k}\sigma} b_{{\bf k}\sigma} + {\rm h.c.} - \frac{ \Delta_a^{2}}{g_a} - \frac{ \Delta_b^{2}}{g_b}.
\end{align}
The operator $a^\dagger_{{\bf k}\sigma}$ creates an electron in band $a$ with momentum ${\bf k}$ and spin $\sigma$, and similarly for band $b$. The kinetic energy is given by the band dispersions $\varepsilon_{\eta}= \frac{k^2}{2m_\eta}-\mu_\eta$, where $\eta=a,b$, with $m_\eta$ denoting the electron mass. $g_a$ and $g_b$ are the coupling constants of the electrons in the respective bands. 

In order to guarantee homogeneous equilibrium, all electrons occupying different bands have the same chemical potential $\mu$. Thus, we can define an effective chemical potential of the electrons in bands $a$ and $b$ as $\mu_\eta=\mu+E_\eta$. The constants $E_\eta$ are the bottom of the specific band $\eta$.

A finite single-particle band scattering, or hybridization, $V({\bf k}) \equiv V_k$, has been considered several times in the literature as, for instance, in~\cite{Japiassu,Moreo,essential}.

The ``original'' (i.e., before the actuation of $V_k$) superconducting intra-band {\it mean-field} order parameters are $\Delta_{a,b}$. It will be assumed here, as usual, spin-singlet, regular $s$-wave superconducting states. However, the methods employed here could be generalized to other kinds of intra-band pairing.

Diagonalization of the kinetic energy, considering a finite $V_k$, gives a Hamiltonian with diagonal bands $c$ and $d$, however now exhibiting both intra-band superconducting order parameters $\Delta_c$ and $\Delta_d$ and inter-band order parameters $\Delta_{cd}$ and $\Delta_{dc}$:
%
\begin{align}
\label{eq:Hcd}
H_{cd}  & =  \sum_{{\bf k}\sigma} \varepsilon_c ({\bf k}) c^\dagger_{{\bf k}\sigma} c_{{\bf k}\sigma} + \varepsilon_d ({\bf k}) d^\dagger_{{\bf k}\sigma} d_{{\bf k}\sigma} \nonumber \\
& + \sum_{\bf k} \Delta_c({\bf k})  c^\dagger_{{\bf k}\uparrow} c^\dagger_{-{\bf k}\downarrow} + \Delta_d({\bf k})  d^\dagger_{{\bf k}\uparrow} d^\dagger_{-{\bf k}\downarrow} +{\rm H.c.} \nonumber \\
& + \sum_{\bf k} \Delta_{cd}({\bf k})  c^\dagger_{{\bf k}\uparrow} d^\dagger_{-{\bf k}\downarrow} + \Delta_{dc}({\bf k})  d^\dagger_{{\bf k}\uparrow} c^\dagger_{-{\bf k}\downarrow} + {\rm H.c.},
\end{align}
where $\varepsilon_{c,d}({\bf k})=\varepsilon_{+}({\bf k}) \pm \sqrt{\varepsilon_{-}({\bf k})^2+ |V_{{\bf k}}|^2}$ and  $\varepsilon_{\pm}({\bf k}) \equiv \frac{\varepsilon_a({\bf k}) \pm \varepsilon_b({\bf k})}{2}$. In the above equation, we have defined

%
\begin{align}
\label{eq:gap-c}
\Delta_{c}({\bf k}) = \frac{A^2 \Delta_{a}({\bf k})  + V_{{\bf k}} V_{-{\bf k}} \Delta_{b}({\bf k}) }{A \sqrt{(\varepsilon_{b}({\bf k})-\varepsilon_{a}({\bf k}))^2+ 4|V_{_{{\bf k}}}|^2}},
\end{align}
where $A\equiv \varepsilon_{-}({\bf k})+ \sqrt{\varepsilon_{-}({\bf k})^2+ |V_{{\bf k}}|^2}$,

%
\begin{align}
\label{eq:gap-d}
\Delta_{d}({\bf k}) = \frac{A^2 \Delta_{b}({\bf k})  +  V_{{\bf k}}^{*} V_{- {\bf k}}^{*} \Delta_{a}({\bf k})}{A \sqrt{(\varepsilon_{b}({\bf k})-\varepsilon_{a}({\bf k}))^2+ 4|V_{{\bf k}}|^2}},
\end{align}

%
\begin{align}
\label{eq:gap-cd}
\Delta_{cd}({\bf k}) = \frac{\Delta_{b}({\bf k}) V_{{\bf k}} - \Delta_{a}({\bf k}) V_{- {\bf k}}^{*}}{\sqrt{(\varepsilon_{b}({\bf k})-\varepsilon_{a}({\bf k}))^2+ 4|V_{{\bf k}}|^2}},
\end{align}

%
\begin{align}
\label{eq:gap-dc}
\Delta_{dc}({\bf k}) = \frac{\Delta_{b}({\bf k}) V_{- {\bf k}} - \Delta_{a}({\bf k}) V_{{\bf k}}^{*}}{\sqrt{(\varepsilon_{b}({\bf k})-\varepsilon_{a}({\bf k}))^2+ 4|V_{{\bf k}}|^2}}.
\end{align}

Notice that $\Delta_c(V_k=0) \to \Delta_a$, $\Delta_d(V_k=0) \to \Delta_b$ while $\Delta_{cd}(V_k=0) \to 0$, and $\Delta_{dc}(V_k=0) \to 0$. Notice also that if we take a symmetric $V_{{\bf k}}$, such that $V_{- {\bf k}} = V_{{\bf k}}$, and $V_{- {\bf k}}^{*}=V_{{\bf k}}$, we have that

%
\begin{align}
\label{eq:Dab}
\Delta_{cd}({\bf k})=\Delta_{dc}({\bf k}) = \frac{V_{{\bf k}} (\Delta_{b}({\bf k}) - \Delta_{a}({\bf k}) )}{\sqrt{(\varepsilon_{b}({\bf k})-\varepsilon_{a}({\bf k}))^2+ 4|V_{{\bf k}}|^2}}.
\end{align}
This gap corresponds to the one derived in \cite{essential}.

However, an antisymmetric hybridization, $V_{- {\bf k}} = -V_{{\bf k}}$, gives 
$$\Delta_{cd}({\bf k}) = \frac{\Delta_{b}({\bf k}) V_{{\bf k}} + \Delta_{a}({\bf k}) V_{{\bf k}}^{*}}{\sqrt{(\varepsilon_{b}({\bf k})-\varepsilon_{a}({\bf k}))^2+ 4|V_{{\bf k}}|^2}}$$ 
and 
$$\Delta_{dc}({\bf k}) = -\frac{\Delta_{b}({\bf k}) V_{{\bf k}} + \Delta_{a}({\bf k}) V_{{\bf k}}^{*}}{\sqrt{(\varepsilon_{b}({\bf k})-\varepsilon_{a}({\bf k}))^2+ 4|V_{{\bf k}}|^2}}.$$ 
Besides being antisymmetric, a pure imaginary hybridization~\cite{Annals1} $V_{{\bf k}}= i \gamma ({\bf k_x} k_x + {\bf k_y} k_y + {\bf k_z} k_z)$ (such the one that arises in one-dimension), where $\gamma$ again denotes the strength of the hybridization, implies in

%
\begin{align}
\label{eq:Dcd}
\Delta_{cd}({\bf k}) = \frac{V_{{\bf k}} (\Delta_{b}({\bf k}) - \Delta_{a}({\bf k}) )}{2\sqrt{\delta \varepsilon^2+ |V_{{\bf k}}|^2}} = - \Delta_{dc}({\bf k}).
\end{align}
where we defined $\delta \varepsilon= \frac{\varepsilon_{b}({\bf k})-\varepsilon_{a}({\bf k})}{2}=\frac{\mu_a-\mu_b}{2}=\frac{E_a-E_b}{2}$, since we consider the same mass for the electrons in the bands $a$ and $b$, $m_a=m_b$. 

As we pointed out earlier, the (induced) inter-band pairing gaps vanish for $V_{{\bf k}} = 0$ (i.e., in the absence of hybridization), in which case the bands behave as completely two independent ones. Besides, it is also intriguing to note that this induced {\it inter-band} pairing gap will not emerge in the case the gaps in the non-interacting bands have the same magnitude i.e., $\Delta_{a}({\bf k}) =\Delta_{b}({\bf k})$, no matter how strong the strength the hybridization is. It is also worth to observe that the pairing gaps in Eq.~(\ref{eq:Dcd}) are of $p$-wave type, due to the antisymmetric character of $V_{{\bf k}}$. 

The gaps in Eq.~(\ref{eq:Dcd}) should be compared with similar p-wave pairing gaps derived in the setups proposed in \cite{Sau}, where a semiconductor quantum well coupled to an s-wave superconductor and a ferromagnetic insulator, and an alternative one~\cite{Alicea}, in which a topological superconducting phase is driven by applying an in-plane magnetic field to a $110$- grown semiconductor coupled only to an s-wave superconductor.

In order to make a more concrete comparison between the (induced) p-wave gaps, the one obtained here resulting from the hybridization of the $a$ and $b$ bands, and the one obtained in the one-band model with spin-orbit (SO) coupling, but with a magnetic field raising the degenerescence between the up and down electrons~\cite{Sau,Alicea}, we drop one of the bands, say the $b$-band for instance. Since we are in a one band problem now, we identify $\delta \varepsilon^2$ as an (external) Zeeman magnetic field $V_z$ responsible for raising the degeneracy between the spin-up and spin-down pairing species, to obtain

\begin{align}
\label{eq:1band}
\Delta_{cd}({\bf k}) = \frac{-V_{k}  \Delta }{2\sqrt{|V_k|^2+ V_z^2}} = - \Delta_{dc}({\bf k}),
\end{align}
which is the expression for the (intra-band) induced by SO coupling p-wave pairing gap derived in~\cite{Alicea}.

\section{\label{gap} Gap Equations with Symmetric Hybridization}

The Hamiltonian in Eq.~(\ref{eq:Hab}) can be rewritten in the basis~$\Psi_{{\bf k}} = (a_{\textbf{k},\uparrow}, b_{\textbf{k}, \uparrow},
a_{-\textbf{k},\downarrow}^{\dag}, b_{-\textbf{k},\downarrow}^{\dag})^{T}$ as:

\begin{eqnarray}
H=\frac{1}{2} \sum_{{\bf k}}\Psi_{{\bf k}}^{\dag}\mathcal{H}({\bf k})\Psi_{{\bf k}}
 + 2 \sum_{\textbf{k}} \varepsilon_{+}({\bf k}) - \frac{ \Delta_a^{2}}{g_a} - \frac{ \Delta_b^{2}}{g_b}, 
 \label{6}
\end{eqnarray}
with

\begin{eqnarray}
\mathcal{H}({\bf k})=\left(
  \begin{array}{cccc}
    \varepsilon_a ({\bf k}) & V_{{\bf{k}}}^{*}  & -\Delta_{a}^{*} &  0  \\
 V_{{\bf{k}}} & \varepsilon_b ({\bf k}) & 0 & -\Delta_{b}^{*} \\
   - \Delta_a & 0 & -\varepsilon_a ({\bf k}) &  -V_{{\bf{k}}} \\
    0 & -\Delta_b & -V_{{\bf{k}}}^{*} &  -\varepsilon_b ({\bf k})  \\
  \end{array}
\right). \label{7}
\end{eqnarray}
This Hamiltonian can be diagonalized as
\begin{eqnarray}
H&=&\sum_{\textbf{k},s=1,2} E_{\textbf{k},s}\alpha_{\textbf{k},s}^{\dag}\alpha_{\textbf{k},s} \\
\nonumber
&+& \sum_{\textbf{k},s=1,2} (2\varepsilon_{+}({\bf k}) -E_{\textbf{k},s}) - \frac{\Delta_a^{2}}{g_a} - \frac{\Delta_b^{2}}{g_b}, 
\label{8}
\end{eqnarray}
where, $\alpha_{\textbf{k}, 1,2}^{\dag}(\alpha_{\textbf{k}, 1,2})$ is the creation (annihilation) operator for the quasiparticles with excitation spectra

\begin{widetext}
\begin{eqnarray}
E_{{\bf{k}}, 1,2}\!=\!\frac{1}{2} \sqrt{ 2  E_{{\bf{k}}}^{2} \!+\! 4 |V_{{\bf{k}}}|^2 \!\pm \!2 \sqrt{\! \big( |\Delta_a|^{2}\!-\!|\Delta_b|^{2} \!+\! \varepsilon_a ({\bf k})^2 \!-\! \varepsilon_b ({\bf k})^2 \big)^2\!+\! 4|V_{{\bf{k}}}|^2\Big[\big(\big(\varepsilon_a ({\bf k}) \!+\! \varepsilon_b ({\bf k}) \big)^2 \!+\! |\Delta_{a}|^{2} \!+\! |\Delta_{b}|^{2}\Big] \!-\! 8 Re[\Delta_{a}\Delta_{b}^{*} V_{{\bf{k}}}^{2}]} }, \nonumber\\
\end{eqnarray}
where we have defined $E_{{\bf{k}}}^{2} \equiv  |\Delta_a|^{2}+|\Delta_b|^{2} + \varepsilon_a ({\bf k})^2 + \varepsilon_b ({\bf k})^2 $. The other two quasiparticles energies are $E_{{\bf{k}}, 3}=-E_{{\bf{k}}, 1}$ and $E_{{\bf{k}}, 4}=-E_{{\bf{k}}, 2}$.  If we take the order parameters and the hybridization as real terms, the previous equation can be simplified as
\begin{eqnarray}
E_{{\bf{k}}, 1,2}=\frac{1}{2} \sqrt{ 2  E_{{\bf{k}}}^{2} + 4 V_{{\bf{k}}}^2 \pm 2 \sqrt{ \big( \Delta_a^{2}-\Delta_b^{2} + \varepsilon_a ({\bf k})^2 - \varepsilon_b ({\bf k})^2 \big)^2+ 4V_{{\bf{k}}}^2\Big[\big(\big(\varepsilon_a ({\bf k}) + \varepsilon_b ({\bf k}) \big)^2 + \big(\Delta_{a} - \Delta_{b}\big)^{2} \Big] } }. \nonumber\\
\label{Energias}
\end{eqnarray}
\end{widetext}
 It is worth to notice that $\Delta_a^2+\Delta_b^2$ present in $E_{{\bf{k}}, 1,2}$ can be expressed in terms of the induced pairing gap obtained in the previous section as $\Delta_a^2+\Delta_b^2 = \Delta_c^2+\Delta_d^2 + 2\Delta_{cd}^2$, where we have used Eqs.~(\ref{eq:gap-c}) to (\ref{eq:gap-cd}). This shows that if one diagonalizes the full Hamiltonian, as we did here in this section, one does not see that there is an induced gap embedded in the results.

It is very easy to verify that when $V_k=0$ we obtain $E_{\textbf{k}, 1} = \sqrt{ \varepsilon_a ({\bf k})^2 + \Delta_a^{2}} $ and $E_{\textbf{k},2}= \sqrt{ \varepsilon_b ({\bf k})^2 + \Delta_b^{2}} $, which are the quasiparticle dispersions of two independent BCS superconductors, as it should be.

In the case of only interband interaction $V_k$, with $\Delta_a=\Delta_b=0$, the quasiparticles energies are written as

\begin{widetext}
\begin{eqnarray}
E_{{\bf{k}}, 1,2}&=&\frac{1}{2} \sqrt{ 2  (\varepsilon_a({\bf k})^2+\varepsilon_b({\bf k})^2) + 4 V_{{\bf{k}}}^2 \pm 2 \sqrt{\big(  \varepsilon_a ({\bf k})^2 - \varepsilon_b ({\bf k})^2\big)^2 + 4V_{{\bf{k}}}^2\left[ \big( (\varepsilon_a ({\bf k}) + \varepsilon_b ({\bf k}) \big)^2\right]} }\\
\nonumber
&=& \varepsilon_{+} ({\bf k}) \pm \sqrt{ \varepsilon_{-} ({\bf k})^2 + V_{{\bf{k}}}^2},
\label{Energias2}
\end{eqnarray}
\end{widetext}
where $\varepsilon_{\pm}({\bf k}) \equiv \frac{\varepsilon_a({\bf k}) \pm \varepsilon_b({\bf k})}{2}$.

Another immediate result is that of bands with equal gap parameters, $\Delta_a = \Delta_b=\Delta$, in which case, from Eq.~(\ref{Energias}) one finds

\begin{eqnarray}
E_{{\bf{k}}, 1,2}= \sqrt{ \Delta^{2} + \left( \varepsilon_{+} ({\bf k}) \pm \sqrt{ \varepsilon_{-} ({\bf k})^2 + V_{{\bf{k}}}^2} \right)^2 }.
\end{eqnarray}
Although this case is not of particular interest in this work~\footnote{As we have seen in the previous subsection, bands having gap parameters with the same magnitude do not allow the emergence of (induced) p-wave pairing gaps, which is our main aim here.}, for the sake of completeness we will obtain below the gap equation and critical temperature for such a system.

It is straightforward to write down the grand thermodynamic potential $\Omega = -  \text{Tr}\ln[e^{-\beta H}]$, where $\beta=1/(k_{B} T)$, at finite temperature,

\begin{eqnarray}
\label{poteff1}
\Omega &=&
\frac{1}{2}\sum_{\textbf{k},s=1,2} \left[\varepsilon_{+}({\bf k})-E_{\textbf{k},s} -\frac{2}{\beta} \ln(1+e^{-\beta E_{\textbf{k},s}}) \right] \nonumber\\
&-& \frac{\Delta_a^{2}}{g_a} - \frac{\Delta_b^{2}}{g_b} . 
\end{eqnarray}
For the case we are considering now, of two hybridized bands with gap parameters with the same magnitude, we have

\begin{eqnarray}
\label{poteff2}
\Omega \!=\!
\frac{1}{2}\sum_{\textbf{k},s=1,2} \left[\varepsilon_{+}({\bf k})\! -\!E_{\textbf{k},s}\! -\! \frac{2}{\beta} \ln(1\!+\!e^{-\beta E_{\textbf{k},s}}) \right] \!-\!  \frac{\Delta^{2}}{g}, \nonumber\\
\end{eqnarray}
from which we obtain the gap equation

\begin{eqnarray}
\label{gapequa1}
\frac{1}{g} = \frac{1}{4} \sum_{\textbf{k}} \left[ \frac{\tanh(E_{\textbf{k},1}/2T)}{E_{\textbf{k},1}} +\frac{\tanh(E_{\textbf{k},2}/2T)}{E_{\textbf{k},2}}  \right].
\end{eqnarray}
At the critical temperature $T_c$, $\Delta=0$, and the above equation turns out to be

\begin{eqnarray}
\label{gapequa2}
\frac{1}{g} \! =\! \frac{1}{4}\! \sum_{\textbf{k}} \! \left[ \frac{\tanh(\varepsilon_{1}({\bf k})/2T_c)}{\varepsilon_{1}({\bf k})} \! +\! \frac{\tanh(\varepsilon_{2}({\bf k})/2T_c)}{\varepsilon_{2}({\bf k})}  \right],
\end{eqnarray}
where $\varepsilon_{1,2}({\bf k})=\varepsilon_{+}({\bf k}) \pm \sqrt{\varepsilon_{-}({\bf k})^2+ |V_{{\bf{k}}}|^2}$. To solve Eq.~(\ref{gapequa2}) analytically, we consider the case where $\varepsilon_a({\bf k}) \approx \varepsilon_b({\bf k})$, in which case we obtain $\varepsilon_{1,2}({\bf k}) \approx \varepsilon_{+}({\bf k}) \pm V$, where we consider the hybridization is constant and equal to $V$. With this approximation, the gap equation can be written as

\begin{eqnarray}
\label{gapequa3}
\frac{1}{\lambda} \approx \frac{1}{2} \int_0^{\omega_D} d\varepsilon \left[ \frac{\tanh(\varepsilon-V/2T_c)}{\varepsilon-V} +\frac{\tanh(\varepsilon+V/2T_c)}{\varepsilon+V}  \right], \nonumber\\
\end{eqnarray}
where $\lambda \equiv g \rho(0)$, $\omega_D$ is the Debye frequency and $\rho(0)$ is the density of states at the Fermi level. Then we find

\begin{eqnarray}
\label{gapequa4}
T_c = \frac{2 e^{\gamma}}{\pi} \sqrt{\omega_D^{2} - V^2} ~ e^{-1/\lambda},
\end{eqnarray}
which is valid for $\varepsilon_{-}({\bf k}) \approx 0$ and $V < \omega_D$. This equation reveals the existence of a critical hybridization $V_c=\omega_D$ at which superconductivity is disrupted. When the symmetric (and constant) hybridization between the bands is of the order of the energy scale of the phonons responsible for pairing, superconductivity is completely destroyed~\cite{Ramos}.

Turning now to the general case, minimizing $\Omega$ in Eq.~(\ref{poteff1}) with respect to the gaps $\Delta_a$ and $\Delta_b$ respectively, gives

\begin{eqnarray}
\label{gapequaG1}
\frac{4\Delta_a}{g_a} \!=\! \sum_{\textbf{k}} \! \left[\tanh \left(\frac{E_{\textbf{k},1}}{2T} \right) \frac{\partial E_{\textbf{k},1}}{\partial \Delta_a} \!+\! \tanh\left(\frac{E_{\textbf{k},2}}{2T}\right) \frac{\partial E_{\textbf{k},2}}{\partial \Delta_a}  \right], \nonumber\\
\end{eqnarray}
and

\begin{eqnarray}
\label{gapequaG2}
\frac{4\Delta_b}{g_b} \!=\! \sum_{\textbf{k}} \! \left[\tanh \left(\frac{E_{\textbf{k},1}}{2T} \right) \frac{\partial E_{\textbf{k},1}}{\partial \Delta_b} \!+ \! \tanh \left( \frac{E_{\textbf{k},2}}{2T} \right) \frac{\partial E_{\textbf{k},2}}{\partial \Delta_b}  \right]. \nonumber\\
\end{eqnarray}

From here now we concentrate in obtaining the zero temperature gaps. Eqs~(\ref{gapequaG1}) and (\ref{gapequaG2}) at zero $T$ read,

\begin{eqnarray}
\label{gapequaG1T0}
\frac{4\Delta_a}{g_a} \!=\! \int \frac{d^3 k}{(2 \pi)^3} \left[ \frac{\partial E_{\textbf{k},1}}{\partial \Delta_a} \!+ \frac{\partial E_{\textbf{k},2}}{\partial \Delta_a}  \right],
\end{eqnarray}
and

\begin{eqnarray}
\label{gapequaG2T0}
\frac{4\Delta_b}{g_b} \!=\! \int \frac{d^3 k}{(2 \pi)^3} \left[ \frac{\partial E_{\textbf{k},1}}{\partial \Delta_b} \!+ \frac{\partial E_{\textbf{k},2}}{\partial \Delta_b}  \right].
\end{eqnarray}

In order to integrate Eqs.~(\ref{gapeqT0}) we verified that after some simple algebra (see (\ref{Ap1}) to (\ref{Ap3})) the momentum dependent terms in Eq.~(\ref{Energias}) can be written as

\begin{eqnarray}
\label{simple}
\varepsilon_a ({\bf k}) + \varepsilon_b ({\bf k}) &=& 2 {\xi_k}\\
\nonumber
\varepsilon_a ({\bf k})^2 - \varepsilon_b ({\bf k})^2 &=& 4 \delta \mu \xi_k,\\
\nonumber
\varepsilon_a ({\bf k})^2 + \varepsilon_b ({\bf k})^2 &=& 2 [{\xi_k}^2+{\delta \mu}^2],
\nonumber
\end{eqnarray}
where we have defined $\xi_k = \frac{k^2}{2m} - \bar \mu $, $\bar \mu = (\mu_a + \mu_b)/2=(2 \mu + E_a + E_b)/2$ and $\delta \mu = (\mu_b - \mu_a)/2 $. This will allow us to make the appropriate (usual) change of variables $\xi = \frac{k^2}{2m} - \bar \mu $ to proceed with the integration of the gap equations in Eqs.~(\ref{gapeqT0}),

\begin{widetext}
\begin{eqnarray}
\label{gapeqT02}
\frac{\Delta_a}{\lambda_a} \!=\! \frac{1}{4} \int_0^{\omega} d \xi \left[ \frac{\Delta_a}{E_{\xi,1}} \!+ \frac{\Delta_a}{E_{\xi,2}}  + \frac{1}{E(\xi)} \left[ \Delta_a(\Delta_a^{2}-\Delta_b^{2} +4 \delta \mu \xi ) + 2V_{\xi}^2 \left(\Delta_a-\Delta_b \right) \right] \left[ \frac{1}{E_{\xi,1}} \! - \frac{1}{E_{\xi,2}} \!  \right] \!  \right],
\\
\nonumber
\frac{\Delta_b}{\lambda_b} \!=\! \frac{1}{4} \int_0^{\omega} d \xi \left[ \frac{\Delta_b}{E_{\xi,1}} \!+ \frac{\Delta_b}{E_{\xi,2}}  - \frac{1}{E(\xi)} \left[ \Delta_b(\Delta_a^{2}-\Delta_b^{2} +4 \delta \mu \xi ) + 2V_{\xi}^2 \left(\Delta_a-\Delta_b \right) \right] \left[ \frac{1}{E_{\xi,1}} \! - \frac{1}{E_{\xi,2}} \!  \right] \!  \right],
\end{eqnarray}
\end{widetext}
where $\lambda_a \equiv g_a \rho(0)$, $\lambda_b \equiv g_b \rho(0)$, with $\rho(0) = \frac{m}{2 \pi^2} k_F$ and $\omega$ is an energy cutoff. Here $k_F=\sqrt{2m \bar \mu}$ is the Fermi momentum.  In the above equations we have taken a symmetric hybridization $V_k = \gamma k^2$, where $\gamma$ is the strength of the hybridization, and $k^2 = k_x^2+ k_y^2+ k_z^2$, such that $V(-k)=V(k)$. Thus, $V_{\xi} = 2m \gamma (\bar \mu + \xi)$. The ``quasiparticle energies'' now read

\begin{widetext}
\begin{eqnarray}
\label{EnergiasNew}
E_{{\xi}, 1,2}=\frac{1}{2} \sqrt{ 2  E_{\xi}^{2} + 4 V_{\xi}^2 \pm 2 \sqrt{ \big( \Delta_a^{2}-\Delta_b^{2} + 4 \delta \mu \xi \big)^2+ 4V_{\xi}^2\Big[\big(\big( 2 \xi \big)^2 + \big(\Delta_{a} - \Delta_{b}\big)^{2} \Big] } }. \nonumber\\
\end{eqnarray}
\end{widetext}
Besides, $E_{\xi}^{2} =  \Delta_a^{2}+\Delta_b^{2} + 2 [{\xi}^2+{\delta \mu}^2] $, and $E(\xi) = \Big\{\left(\Delta_a^{2}-\Delta_b^{2} + 4 \delta \mu \xi \right)^2 + 4V_{\xi}^2\left[\left(\Delta_a-\Delta_b \right)^{2} + \left( 2 \xi \right)^2\right] \Big\}^{1/2}$. Notice that Eqs.~(\ref{gapeqT02}) as well as Eqs.~(\ref{EnergiasNew}) are symmetric under the (simultaneous) transformation $\Delta_a \to -\Delta_a$ {\it and} $\Delta_b \to -\Delta_b$. 

Notice also that the trivial solutions $\Delta_a=0$ and $\Delta_b=0$ are solutions of the gap equations in Eqs.~(\ref{gapeqT02}). Taking $\Delta_a=0$ in the first equation, and $\Delta_b=0$ in the second one, these equations give, respectively 

\begin{eqnarray}
0=\int_0^{\omega} d \xi~\frac{V_{\xi}^2 \Delta_b}{E(\xi)} \left[ \frac{1}{E_{\xi,1}} \! - \frac{1}{E_{\xi,2}} \!  \right],\\
\nonumber
0=\int_0^{\omega} d \xi~\frac{V_{\xi}^2 \Delta_a}{E(\xi)} \left[ \frac{1}{E_{\xi,1}} \! - \frac{1}{E_{\xi,2}} \!  \right].
\end{eqnarray}
Since $V_{\xi}^2 \neq 0$, and $E_{\xi, 1}(\Delta_a=\Delta_b=0) = | \xi + \sqrt{ \tilde \delta \mu^2 + V_\xi^2}| \neq E_{\xi, 2}(\Delta_a=\Delta_b=0) = | \xi - \sqrt{ \tilde \delta \mu^2 + V_\xi^2}|$, the only solution is $\Delta_a=0$ and $\Delta_b=0$. This fact will be evident below, in the numerical evaluation of the gap equations as a function of the hybridization.

Defining now the non-dimensional variables $x=\xi/E_F$, $\tilde \Delta_{a,b} = \Delta_{a,b}/E_F$, $\tilde \delta \mu = \delta \mu/E_F$, $\tilde \mu = \bar \mu/E_F$, and $\tilde V_x = V_\xi/E_F = 2m \gamma (\tilde \mu +x)$, where $E_F = k_F^2/2m$ is the Fermi energy. Defining also a non-dimensional (for the case of the symmetric hybridization we are considering) hybridization parameter $\alpha \equiv 2 m \gamma$, we have $\tilde V_x^2 = \alpha^2 (\tilde \mu +x)^2$. Thus, the gap equations in (\ref{gapeqT02}) are rewritten as

\begin{widetext}
\begin{eqnarray}
\label{gapeqT0x}
\frac{\Delta_a}{\lambda_a} \!=\! \frac{1}{4} \int_0^{X} d x \left[ \frac{\Delta_a}{E_{x,1}} \!+ \frac{\Delta_a}{E_{x,2}}  + \frac{1}{E(x)} \left[ \Delta_a(\Delta_a^{2}-\Delta_b^{2} +4 \delta \mu x ) + 2\tilde V_{x}^2 \left(\Delta_a-\Delta_b \right) \right] \left[ \frac{1}{E_{x,1}} \! - \frac{1}{E_{x,2}} \!  \right] \!  \right],
\\
\nonumber
\frac{\Delta_b}{\lambda_b} \!=\! \frac{1}{4} \int_0^{X} d x \left[ \frac{\Delta_b}{E_{x,1}} \!+ \frac{\Delta_b}{E_{x,2}}  - \frac{1}{E(x)} \left[ \Delta_b(\Delta_a^{2}-\Delta_b^{2} +4 \delta \mu \xi ) + 2\tilde V_{x}^2 \left(\Delta_a-\Delta_b \right) \right] \left[ \frac{1}{E_{x,1}} \! - \frac{1}{E_{x,2}} \!  \right] \!  \right],
\end{eqnarray}
\end{widetext}
where $E_{{x}, 1,2}=\frac{1}{2} \sqrt{ 2  E_{x}^{2} + 4 \tilde V_{x}^2 \pm 2  E(x)}$, $E_{x}^{2} =  \Delta_a^{2}+\Delta_b^{2} + 2 ({x}^2+{\delta \mu}^2)$ and $E(x)=\sqrt{ \big( \Delta_a^{2}-\Delta_b^{2} + 4 \delta \mu x \big)^2+ 4 \tilde V_{x}^2\Big[( 2 x )^2 + (\Delta_{a} - \Delta_{b})^{2} \Big] }$. Given $g_a$ and $g_b$, Eqs.~(\ref{gapeqT0x}) have to be solved self-consistently to find the gaps $\Delta_a$ and $\Delta_b$.

\begin{figure}
\centering
 \includegraphics[width=8cm]{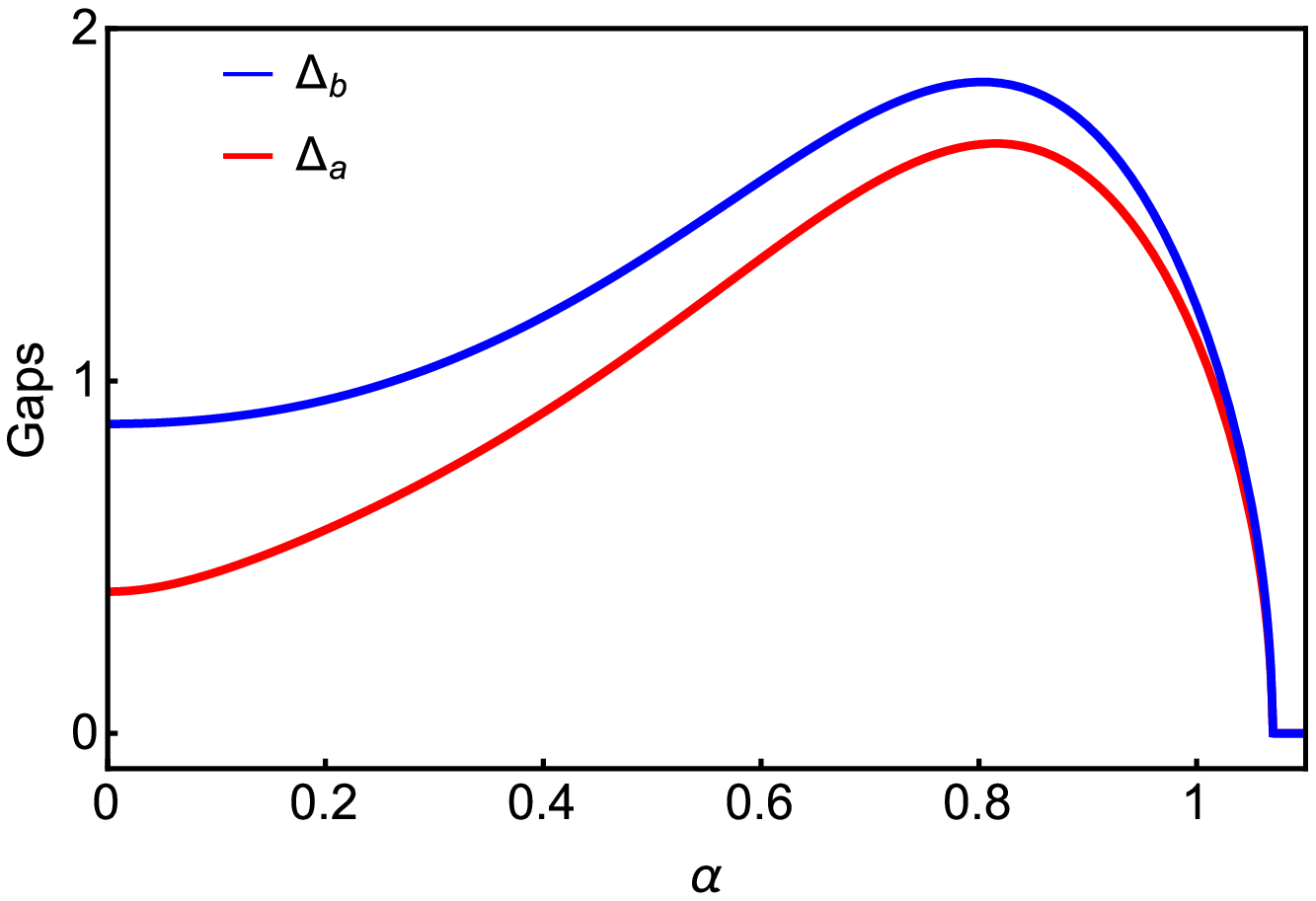}
\caption{(Color online) Gap parameters with symmetric hybridization as a function of $\alpha=2m\gamma$ for $X=10$, $\lambda_a = 0.58$, $\lambda_b = 0.6$, $\tilde \mu_a = 1.2$, and $\tilde \mu_b = 1.6.$}
\label{fig1}
\end{figure}

In Fig.~(\ref{fig1}) we show the gaps $a$ and $b$ as a function of $\alpha$. The curves are the self-consistent solutions of Eq.~(\ref{gapeqT02}). From these curves we can also see that there exists a critical hybridization at which the two gaps vanish.

Notice that as $\alpha$ approaches its critical value, $\Delta_a \approx \Delta_b \equiv \Delta(\alpha)$, where $\Delta(\alpha)$ is the solution of the equation

\begin{equation}
\label{DCrit1}
\frac{1}{\lambda}= \frac{1}{2} \int_0^{\omega} d \xi \left( \frac{1}{E_{\xi,1}} + \frac{1}{E_{\xi,2}} \right),
\end{equation} 
where
$E_{\xi,1,2}= \sqrt{(V_{\xi} \pm \xi)^2+\Delta^2}$, and $V_{\xi} = \alpha (\bar \mu + \xi)$. The integration is straightforward and yields 

\begin{eqnarray}
\label{DCrit2}
\frac{2}{\lambda}&=& \frac{1}{\alpha+1} \ln\left[\frac{2(\alpha+1)\omega}{\sqrt{(\alpha^2 {\bar \mu}^2+\Delta^2}} \right] \\
\nonumber
&+& \frac{1}{|\alpha-1|} \ln\left[\frac{2 |\alpha-1| \omega}{\sqrt{(\alpha^2 {\bar \mu}^2+\Delta^2}} \right].
\end{eqnarray} 
The limit $\alpha \to 0$ gives the well known standard BCS result $\Delta(0)= 2\omega e^{-1/\lambda}$, as expected.

It is also instructive to find the gap which is solution of Eq.~(\ref{DCrit1}) for the case of two bands hybridized by a constant hybridization $V_\xi = V$,

\begin{equation}
\label{DCrit3}
\Delta(V) = 2 \sqrt{\omega^2-V^2} e^{-1/\lambda}.
\end{equation} 
As we found earlier, this equation also reveals a critical value for a constant hybridization between the bands, at which superconductivity vanishes, that is the energy cutoff, namely $V_c = \omega$.

\begin{figure}
\centering
 \includegraphics[width=8cm]{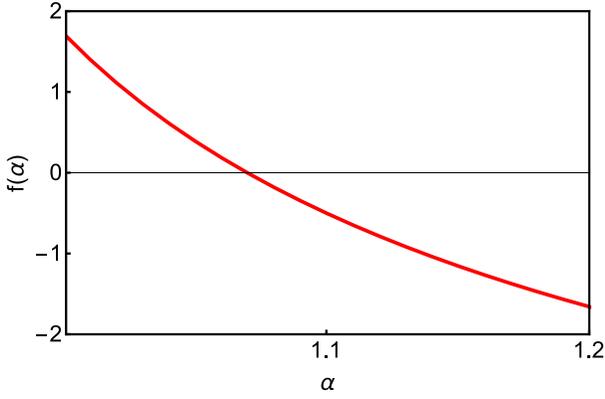}
\caption{Function $f(\alpha)$ as a function of $\alpha$ plotted with the same parameters as in Fig.~\ref{fig1}.}
\label{fig1-2}
\end{figure}

Equation (\ref{DCrit1}) also gives $\alpha_c$, the critical $\alpha$ at which $\Delta$ vanishes, which is the solution of

\begin{equation}
\label{DCrit4}
\frac{1}{\lambda}= \frac{1}{2} \int_0^{\omega} d \xi \left( \frac{1}{V_{\xi}(\alpha_c) + \xi} + \frac{1}{|V_{\xi}(\alpha_c) - \xi|} \right),
\end{equation} 
where $V_{\xi}(\alpha_c)=\alpha_c (\bar \mu + \xi)$. After integration of the above equation we define now a function $f(\alpha)$ given by

\begin{eqnarray}
\label{DCrit5}
f(\alpha)=&-&\frac{2}{\lambda}+ \frac{1}{\alpha+1} \ln\left(\frac{\alpha \bar \mu + (\alpha+1) \omega}{\alpha \bar \mu} \right)\\ 
\nonumber
&+& \frac{1}{\alpha-1} \ln\left(\frac{\alpha \bar \mu + (\alpha-1) \omega}{\alpha \bar \mu}\right).
\end{eqnarray} 
 The zeros of $f(\alpha)$ are the respective $\alpha_c$. Eq.~(\ref{DCrit1}), and hence (\ref{DCrit5}), are valid close to $\alpha_c$. As seen from Fig.~\ref{fig1} $\alpha_c$ is close to 1. The behavior of $f(\alpha)$ in Fig.~\ref{fig1-2} shows that $\alpha_c \approx 1.08$, which agrees with the value shown in Fig.~\ref{fig1}.

\section{\label{gap} Gap Equations with AntiSymmetric Hybridization}

We can use the same procedure developed in the previous section, but  for antisymmetric hybridization, $V_{-k} = - V_{k}$. The Hamiltonian in Eq.~(\ref{eq:Hab}) can be rewritten in the basis~$\Psi_{{\bf k}} = (a_{\textbf{k},\uparrow}, b_{\textbf{k}, \uparrow},
a_{-\textbf{k},\downarrow}^{\dag}, b_{-\textbf{k},\downarrow}^{\dag})^{T}$ as:

\begin{eqnarray}
H=\frac{1}{2} \sum_{{\bf k}}\Psi_{{\bf k}}^{\dag}\mathcal{H}({\bf k})\Psi_{{\bf k}}
 + 2 \sum_{\textbf{k}} \varepsilon_{+}({\bf k}) - \frac{ \Delta_a^{2}}{g_a} - \frac{ \Delta_b^{2}}{g_b}, 
 \label{6}
\end{eqnarray}
with
\begin{eqnarray}
\mathcal{H}({\bf k})=\left(
  \begin{array}{cccc}
    \varepsilon_a ({\bf k}) & - V_{{\bf{k}}}^{*}  & -\Delta_{a}^{*} &  0  \\
 - V_{{\bf{k}}} & \varepsilon_b ({\bf k}) & 0 & -\Delta_{b}^{*} \\
   - \Delta_a & 0 & -\varepsilon_a ({\bf k}) &  -V_{{\bf{k}}} \\
    0 & -\Delta_b & -V_{{\bf{k}}}^{*} &  -\varepsilon_b ({\bf k})  \\
  \end{array}
\right). 
\end{eqnarray}

Diagonalizing the Hamiltonian (\ref{6}), we can write
\begin{eqnarray}
&&H=\sum_{\textbf{k},s=1,2}
E_{\textbf{k},s}\alpha_{\textbf{k},s}^{\dag}\alpha_{\textbf{k},s}
+ \sum_{\textbf{k},s=1,2}
(2\varepsilon_{+}({\bf k}) -E_{\textbf{k},s}) \nonumber\\
\nonumber\\
&&\;\;\;\;- \frac{\Delta_a^{2}}{g_a} - \frac{\Delta_b^{2}}{g_b}, 
\label{h2}
\end{eqnarray}
where, $\alpha_{\textbf{k}, 1,2}^{\dag}(\alpha_{\textbf{k}, 1,2})$ is the creation (annihilation) operator for the quasiparticles with excitation spectra

\begin{widetext}
\begin{eqnarray}
E_{\textbf{k}, 1,2}\!=\!\frac{1}{2} \sqrt{ 2  E_{{\bf{k}}}^{2} \!+\! 4 |V_{{\bf{k}}}|^2 \!\pm \!2 \sqrt{\! \big( |\Delta_a|^{2}\!-\!|\Delta_b|^{2} \!+\! \varepsilon_a ({\bf k})^2 \!-\! \varepsilon_b ({\bf k})^2 \big)^2\!+\! 4|V_{{\bf{k}}}|^2\Big[\big(\big(\varepsilon_a ({\bf k}) \!+\! \varepsilon_b ({\bf k}) \big)^2 \!+\! |\Delta_{a}|^{2} \!+\! |\Delta_{b}|^{2}\Big] \!-\! 8 Re[\Delta_{a}\Delta_{b}^{*} V_{{\bf{k}}}^{2}]} }, \nonumber\\
\end{eqnarray}
where we have defined $E_{{\bf{k}}}^{2} \equiv  |\Delta_a|^{2}+|\Delta_b|^{2} + \varepsilon_a ({\bf k})^2 + \varepsilon_b ({\bf k})^2 $. The other two quasiparticles energies are $E_{\textbf{k}, 3}=-E_{\textbf{k}, 1}$ and $E_{\textbf{k}, 4}=-E_{\textbf{k}, 2}$.  We assume without loss of generality that the order parameters $\Delta_{a}$ and $\Delta_{b}$ are real. Since
the antisymmetric hybridization $V_{{\bf{k}}}$ has to be purely imaginary to preserve time reversal symmetry, the term $Re[\Delta_{a}\Delta_{b}^{*} V_{{\bf{k}}}^2]$ turns out to be the same as $-\Delta_a\Delta_b|V_{{\bf{k}}}|^2$. So the previous equation can be simplified as

\begin{eqnarray}
E_{\textbf{k}, 1,2}=\frac{1}{2} \sqrt{ 2  E_{k}^{2} + 4 |V_{k}|^2 \pm 2 \sqrt{ \big( \Delta_a^{2}-\Delta_b^{2} + \varepsilon_a ({\bf k})^2 - \varepsilon_b ({\bf k})^2 \big)^2\!+\! 4|V_{{\bf{k}}}|^2\Big[\big(\varepsilon_a ({\bf k}) +\varepsilon_b ({\bf k}) \big)^2 + (\Delta_{a} + \Delta_{b})^{2}\Big]} }. \nonumber\\
\label{EnergiasAS}
\end{eqnarray}
\end{widetext}

The zero temperature gap equations are given by Eqs.~(\ref{gapequaAS3}) and (\ref{gapequaAS4}). Making the same change of variables, as we did in the case of symmetric hybridization, we obtain $E_{{\xi}, 1,2}=\frac{1}{2} \sqrt{ 2  E_{\xi}^{2} + 4 |V_{\xi}|^2 \pm 2  E(\xi)}$, $E_{\xi}^{2} =  \Delta_a^{2}+\Delta_b^{2} + 2 ({\xi}^2+{\delta \mu}^2)$ and $E(\xi)=\sqrt{ \big( \Delta_a^{2}-\Delta_b^{2} + 4 \delta \mu \xi \big)^2+ 4|V_{\xi}|^2\Big[( 2 \xi )^2 + (\Delta_{a} + \Delta_{b})^{2} \Big] }$.

As we mentioned before, we take a pure imaginary anti-symmetric hybridization $V_{{\bf k}}= i \gamma ({\bf k_x} k_x + {\bf k_y} k_y + {\bf k_z} k_z)$, such that $|V_k|^2 =  \gamma^2 k^2 \to  2m \gamma^2 (\xi + \bar \mu)$. Since $[\gamma]= [k/m]$, we set $[\gamma] = [k_F/m]$, then $\gamma^2 = k_F^2/m^2$ or $m \gamma^2 = k_F^2/m = 2 E_F$. So we can define the non-dimensional hybridization parameter for antisymmetric hybridization $\alpha = 2m\gamma^2/E_F$, such that $|V_k|^2 =  \alpha (\xi + \bar \mu)$. Then, as we did in the case of symmetric hybridization, the gap equations in terms of the non-dimensional variable $x$ are written as

\begin{widetext}
\begin{eqnarray}
\label{AntiS}
\frac{\tilde \Delta_{a}}{\lambda_a} &=& \frac{1}{4} \int_0^X d x  \left[ \frac{\tilde \Delta_{a}}{E_{x, 1}} + \frac{\tilde \Delta_{a}}{E_{x,2} } \! + \! \frac{ \tilde \Delta_{a} ( \tilde \Delta_{a}^{2} - \tilde \Delta_{b}^{2} + 4\tilde \delta \mu x )+2 |\tilde V_{{\bf{x}}}|^{2} (\tilde \Delta_{a} +\tilde \Delta_{b})}{E(x)} \left( \frac{1}{E_{x, 1}}\!- \frac{1}{E_{x, 2}}\right) \right],\\
\nonumber
\frac{\tilde \Delta_{b}}{\lambda_b} &=& \frac{1}{4} \int_0^X d x  \left[ \frac{\tilde \Delta_{b}}{E_{x, 1}} + \frac{\tilde \Delta_{b}}{E_{x,2} } \! - \! \frac{\tilde \Delta_{b} ( \tilde \Delta_{a}^{2} - \tilde \Delta_{b}^{2} + 4\tilde \delta \mu x ) +2 |\tilde V_{{\bf{x}}}|^{2}(\tilde \Delta_{a} + \tilde \Delta_{b})}{E(x)} \left( \frac{1}{E_{x, 1}}\!- \frac{1}{E_{x, 2}}\right) \right],
\end{eqnarray}
\end{widetext}
where $E_{{x}, 1,2}=\frac{1}{2} \sqrt{ 2  E_{x}^{2} + 4 \tilde V_{x}^2 \pm 2  E(x)}$, $E_{x}^{2} =  \Delta_a^{2}+\Delta_b^{2} + 2 ({x}^2+{\delta \mu}^2)$ and $E(x)=\sqrt{ \big( \Delta_a^{2}-\Delta_b^{2} + 4 \delta \mu x \big)^2+ 4 \tilde V_{x}^2\Big[( 2 x )^2 + (\Delta_{a} + \Delta_{b})^{2} \Big] }$, with $\tilde V_x^2 =  \alpha (x + \tilde \mu)$. 

\begin{figure}
\centering
 \includegraphics[width=8cm]{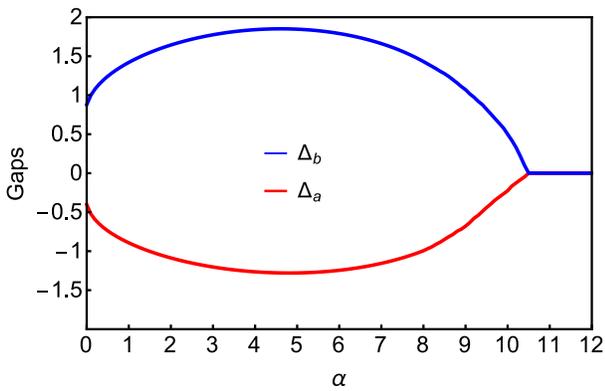}
\caption{(Color online) Gap parameters with antisymmetric hybridization as a function of $\alpha=2m\gamma^2/E_F$ for $X=10$, $\lambda_a = 0.58$, $\lambda_b = 0.6$, $\tilde \mu_a = 1.2$, and $\tilde \mu_b = 1.6.$}
\label{fig2}
\end{figure}

In Fig.~(\ref{fig2}) the two gaps $a$ and $b$ are shown as a function of the hybridization parameter $\alpha=2m\gamma^2/E_F$ for the antisymmetric hybridization we are considering. The curves are the (self-consistent) solutions of Eq.~(\ref{AntiS}).

\section{\label{Co} Summary and Conclusions}

We have investigated the consequences of a symmetric and antisymmetric hybridization $V(k)$ in the superconducting properties of a two-band system. We consider that these bands are formed by electronic orbitals with angular momenta, such that, their hybridization can be symmetric or antisymmetric. We have taken into account only intra-band attractive interactions in the two bands, responsible for (intra-band) $s$-wave pairing gaps $\Delta_a$ and $\Delta_b$, and investigate the appearance of induced inter-band pairing gaps in the system. A symmetric hybridization such that $V(-k)=V(k)$ leads to an induced symmetric inter-band pairing gap $\Delta_{cd}$. As expected, a symmetric $V(k)$ is deleterious for superconductivity, leading to the vanishing of both inter and intra-band gaps as the strength of the hybridization increases and reaches a critical value $\alpha_c$. The induced gap for this symmetric hybridization is small since the gaps approach each other as the hybridization is increased, and vanishes at $\alpha_c$. It is worth to remember here that the induced pairing gaps for symmetric and antisymmetric hybridization depend on the difference $\Delta_b - \Delta_a$.

The scenario is quantitatively and qualitatively changed in the presence of an antisymmetric hybridization satisfying $V(-k)=-V(k)$, which is responsible for the emergence of an induced inter-band $p$-wave pairing gap $\Delta_{cd}$ in the two-band system. In this case the original intra-band gaps gaps $\Delta_a$ and $\Delta_b$ remain relatively far from each other, behaving almost as reflected parabolas. The two gaps will eventually vanish as the strength of the antisymmetric hybridization is increased up to a critical value $\alpha_c$ which is the order of the cutoff energy.

We have also shown that the (interband) superconducting orderings are induced in the total absence of superconducting interaction between the two bands, which is, then, completely dependent on the hybridization between them. This fact makes the hybridization in a two-band system a fundamental ingredient in the obtention of spin-triplet $p$-wave pairing gaps.

Induced $p$-wave pairing gaps became important in the context of Majorana fermions since the models proposed for their experimental realization. The hybridization-induced (HI) spin-triplet $p$-wave pair field $\Delta_{cd}$ studied here avoids the main problem of the SO-induced $p$-wave gaps. The SO p-wave gaps rely on the application of an in-plane magnetic Zeeman field to a semiconductor quantum well coupled to an $s$-wave superconductor. If the Zeeman field is large, compared to the induced gap, the topological phase is destroyed~\cite{Alicea}. Since the HI $p$-wave do not need a Zeeman field, the experimental setup for its realization turns out to be relatively simpler than the SO-induced ones. Based on the fact that hybridization can be tuned by external parameters, such as pressure or doping, the ``production'' of an induced $p$-wave gap with some desired characteristics can, in this way, be controlled.

\section{\label{Ac} Acknowledgments}

HC acknowledges the hospitality of CBPF where part of this work was done. We wish to thank the Brazilian agencies, FAPERJ, CAPES and CNPq for financial support.

\appendix
\section{Change of Variables}
\label{Ap}

Beginning with the single-particle dispersion relations $\varepsilon_{\eta}= \frac{k^2}{2m_\eta}-\mu_\eta$, where $\eta=a,b$, we find

\begin{eqnarray}
\label{Ap1}
\varepsilon_a ({\bf k}) + \varepsilon_b ({\bf k}) = 2\left[ \frac{k^2}{2m}-\bar \mu \right]= 2 {\xi_k}.
\end{eqnarray}

\begin{eqnarray}
\label{Ap2}
\varepsilon_a ({\bf k})^2 - \varepsilon_b ({\bf k})^2 = (\mu_b - \mu_a)(\varepsilon_a ({\bf k}) + \varepsilon_b ({\bf k}))\\
\nonumber
= (\mu_b - \mu_a) \left[ 2 \frac{k^2}{2m} - (\mu_b + \mu_a) \right] = 4 \delta \mu \xi_k.
\end{eqnarray}
Plugging $\varepsilon_a ({\bf k})^2 = \varepsilon_b ({\bf k})^2 + 4 \delta \mu \xi_k$ from the above equation in the equation below, we have

\begin{eqnarray}
\label{Ap3}
\varepsilon_a ({\bf k})^2 + \varepsilon_b ({\bf k})^2\\
\nonumber
= 2\varepsilon_b ({\bf k})^2 + 4 \delta \mu \xi_k = 2 [{\xi_k}^2+{\delta \mu}^2].
\end{eqnarray}

\section{Determination of the Gap Equations}
\label{GE}

\subsubsection{Symmetric Hybridization}

The gap equations are given by

\begin{eqnarray}
\label{gapequaAS1}
\frac{4\Delta_a}{g_a} \!=\! \sum_{\textbf{k}} \! \left[\tanh \left(\frac{E_{\textbf{k},1}}{2T} \right) \frac{\partial E_{\textbf{k},1}}{\partial \Delta_a} \!+\! \tanh\left(\frac{E_{\textbf{k},2}}{2T}\right) \frac{\partial E_{\textbf{k},2}}{\partial \Delta_a}  \right], \nonumber\\
\end{eqnarray}
and

\begin{eqnarray}
\label{gapequaAS2}
\frac{4\Delta_b}{g_b} \!=\! \sum_{\textbf{k}} \! \left[\tanh \left(\frac{E_{\textbf{k},1}}{2T} \right) \frac{\partial E_{\textbf{k},1}}{\partial \Delta_b} \!+ \! \tanh \left( \frac{E_{\textbf{k},2}}{2T} \right) \frac{\partial E_{\textbf{k},2}}{\partial \Delta_b}  \right]. \nonumber\\
\end{eqnarray}
Taking the derivatives of the quasiparticle energies given by Eq.~(\ref{Energias}) with respect to $\Delta_a$ and $\Delta_b$, we obtain

\begin{eqnarray}
\label{gapequaG3}
&&\frac{\partial E_{\textbf{k},1,2}}{\partial \Delta_a} = \frac{1}{2E_{\textbf{k},1,2}} \Bigg[ \Delta_a   \\
\nonumber
&\pm& \left. \frac{1}{E(k)} \Big[ \Delta_a \left(\Delta_a^2 - \Delta_b^2 + \varepsilon_a ({\bf k})^2 - \varepsilon_b ({\bf k})^2 \right) + 2 V_{{\bf{k}}}^2 \left( \Delta_a - \Delta_b \right) \Big] \right],
\end{eqnarray}
and

\begin{eqnarray}
\label{gapequaG4}
&&\frac{\partial E_{\textbf{k},1,2}}{\partial \Delta_b} = \frac{1}{2E_{\textbf{k},1,2}} \Bigg[ \Delta_b \\
\nonumber
&\mp& \left. \frac{1}{E(k)} \Big[ \Delta_b \left( \Delta_a^2 - \Delta_b^2 + \varepsilon_a ({\bf k})^2 - \varepsilon_b ({\bf k})^2 \right) + 2 V_{{\bf{k}}}^2 \left( \Delta_a - \Delta_b \right)\Big] \right],
\end{eqnarray}
where $E(k) \equiv \Big\{\left(\Delta_a^{2}-\Delta_b^{2} + \varepsilon_a ({\bf k})^2 - \varepsilon_b ({\bf k})^2 \right)^2 + 4V_{{\bf{k}}}^2\left[\left(\Delta_a-\Delta_b \right)^{2} + \left(\varepsilon_a ({\bf k}) + \varepsilon_b ({\bf k})\right)^2\right] \Big\}^{1/2} = E_{\textbf{k},1}^2 - E_{\textbf{k},2}^2$.

Substituting the partial derivatives derived in Eqs~(\ref{gapequaG3}) and (\ref{gapequaG4}) in the gap equations, and integrating over the angles, yields

\begin{widetext}
\begin{eqnarray}
\label{gapeqT0}
\frac{\Delta_a}{g_a} \!=\! \frac{1}{8} \int \frac{d k~k^2}{2 \pi^2} \left[ \frac{\Delta_a}{E_{\textbf{k},1}} \!+ \frac{\Delta_a}{E_{\textbf{k},2}}   + \frac{1}{E(k)} \left[ \Delta_a(\Delta_a^{2}-\Delta_b^{2} + \varepsilon_a ({\bf k})^2 - \varepsilon_b ({\bf k})^2 ) + 2V_{{\bf{k}}}^2 \left(\Delta_a-\Delta_b \right) \right] \left[ \frac{1}{E_{\textbf{k},1}} \! - \frac{1}{E_{\textbf{k},2}} \!  \right] \!  \right],
\\
\nonumber
\frac{\Delta_b}{g_b} \!=\! \frac{1}{8} \int \frac{d k~k^2}{2 \pi^2} \left[ \frac{\Delta_b}{E_{\textbf{k},1}} \!+ \frac{\Delta_b}{E_{\textbf{k},2}}  - \frac{1}{E(k)} \left[ \Delta_b(\Delta_a^{2}-\Delta_b^{2} + \varepsilon_a ({\bf k})^2 - \varepsilon_b ({\bf k})^2)  + 2V_{{\bf{k}}}^2 \left(\Delta_a-\Delta_b \right) \right] \left[ \frac{1}{E_{\textbf{k},1}} \! - \frac{1}{E_{\textbf{k},2}} \!  \right] \!  \right].
\end{eqnarray}
\end{widetext}

Notice that from the equations above for $V_k=0$ we obtain the particular case,

\begin{eqnarray}
\label{gapequationsT02}
\frac{1}{g_a} = \frac{1}{4} \int \frac{d^3 k}{(2 \pi)^3}  \frac{1}{E_{\textbf{k},1}},\\
\nonumber
\frac{1}{g_b} =\frac{1}{4} \int \frac{d^3 k}{(2 \pi)^3}  \frac{1}{E_{\textbf{k},2}},
\end{eqnarray}
where $E_{\textbf{k}, 1,2} = \sqrt{ \varepsilon_{a,b} ({\bf k})^2 + \Delta_{a,b}^{2}} $. These are, of course, the gap equations of two independent ``conventional superconductors'' of BCS type. Observe also that in order to have the strict BCS result, besides setting $\varepsilon_{a} ({\bf k})=\varepsilon_{b} ({\bf k})$, and $\Delta_{a}=\Delta_{b}$ in the above equations, in light of Eq.~(\ref{poteff1}) to Eq.~(\ref{gapequa3}) we have to add these two equations, respecting $1/g_a+1/g_b=1/g$.

\subsubsection{Antisymmetric Hybridization}

The gap equations, as before, are

\begin{eqnarray}
\label{gapequaAS3}
\frac{4\Delta_a}{g_a} \!=\! \sum_{\textbf{k}} \! \left[\tanh \left(\frac{E_{\textbf{k},1}}{2T} \right) \frac{\partial E_{\textbf{k},1}}{\partial \Delta_a} \!+\! \tanh\left(\frac{E_{\textbf{k},2}}{2T}\right) \frac{\partial E_{\textbf{k},2}}{\partial \Delta_a}  \right], \nonumber\\
\end{eqnarray}
and

\begin{eqnarray}
\label{gapequaAS4}
\frac{4\Delta_b}{g_b} \!=\! \sum_{\textbf{k}} \! \left[\tanh \left(\frac{E_{\textbf{k},1}}{2T} \right) \frac{\partial E_{\textbf{k},1}}{\partial \Delta_b} \!+ \! \tanh \left( \frac{E_{\textbf{k},2}}{2T} \right) \frac{\partial E_{\textbf{k},2}}{\partial \Delta_b}  \right], \nonumber\\
\end{eqnarray}
where, with the quasiparticle energies given by Eq.~(\ref{EnergiasAS}), we obtain

\begin{eqnarray}
\label{gapequaAS5}
&&\frac{\partial E_{\textbf{k},1,2}}{\partial \Delta_a} = \frac{1}{2E_{\textbf{k},1,2}} \Bigg[ \Delta_a   \\
\nonumber
&\pm& \left. \frac{1}{E(k)} \Big[  \Delta_a(\Delta_a^2 - \Delta_b^2 + \varepsilon_a ({\bf k})^2 - \varepsilon_b ({\bf k})^2)  + 2 |V_{{\bf{k}}}|^2 (\Delta_{a} + \Delta_{b})    \Big] \right],
\end{eqnarray}
and

\begin{eqnarray}
\label{gapequaAS6}
&&\frac{\partial E_{\textbf{k},1,2}}{\partial \Delta_b} = \frac{1}{2E_{\textbf{k},1,2}} \Bigg[ \Delta_b \\
\nonumber
&\mp& \left. \frac{1}{E(k)} \Big[  \Delta_b(\Delta_a^2 - \Delta_b^2 + \varepsilon_a ({\bf k})^2 - \varepsilon_b ({\bf k})^2) + 2 |V_{{\bf{k}}}|^2 (\Delta_{a} + \Delta_{b})  \Big] \right],
\end{eqnarray}
with $E(k) \equiv \Big\{\left(\Delta_a^{2}-\Delta_b^{2} + \varepsilon_a ({\bf k})^2 - \varepsilon_b ({\bf k})^2 \right)^2 + 4|V_{{\bf{k}}}|^2\left[ \left((\varepsilon_a ({\bf k}) + \varepsilon_b ({\bf k})\right)^2 + (\Delta_a +\Delta_b)^{2} \right] \Big\}^{1/2} = E_{\textbf{k},1}^2 - E_{\textbf{k},2}^2$.
Thus we have

\begin{widetext}
\begin{eqnarray}
\frac{4 \Delta_{a}}{g_{a}} \!=\! \frac{1}{2} \sum_{{\bf{k}}} \big[ \frac{\Delta_{a} \tanh( \frac{\beta E_{\textbf{k}, 1}}{2})}{E_{\textbf{k}, 1}} + \frac{\Delta_{a} \tanh (\frac{\beta E_{\textbf{k}, 2}}{2})}{E_{\textbf{k}, 2}}\\
 \nonumber
 +  \frac{\Delta_{a} (\Delta_{a}^2 - \Delta_{b}^2+ \varepsilon_{a}({\bf k})^{2} - \varepsilon_{b}({\bf k})^{2}) + 2 |V_{{\bf{k}}}|^{2}(\Delta_{a} + \Delta_{b})}{E_{\textbf{k}, 1}^{2} - E_{\textbf{k}, 2}^{2}} \left( \frac{\tanh (\frac{\beta E_{\textbf{k}, 1}}{2})}{E_{\textbf{k}, 1}}\!-\! \frac{\tanh (\frac{\beta E_{\textbf{k}, 2}}{2})}{E_{\textbf{k}, 2}}\right) \big], 
\nonumber\\
\frac{4 \Delta_{b}}{g_{b}} \!=\!\frac{1}{2} \sum_{{\bf{k}}} \big[ \frac{\Delta_{b} \tanh( \frac{\beta E_{\textbf{k}, 1}}{2})}{E_{\textbf{k}, 1}} + \frac{\Delta_{b} \tanh (\frac{\beta E_{\textbf{k}, 2}}{2})}{E_{\textbf{k}, 2}} \\
\nonumber
 - \frac{\Delta_{b}(\Delta_{a}^2 - \Delta_{b}^2+ \varepsilon_{a}({\bf k})^{2} - \varepsilon_{b}({\bf k})^{2}) + 2 |V_{{\bf{k}}}|^{2}(\Delta_{a} + \Delta_{b})}{E_{\textbf{k}, 1}^{2} - E_{\textbf{k}, 2}^{2}} \left( \frac{\tanh (\frac{\beta E_{\textbf{k}, 1}}{2})}{E_{\textbf{k}, 1}}\!-\! \frac{\tanh (\frac{\beta E_{\textbf{k}, 2}}{2})}{E_{\textbf{k}, 2}}\right) \big].
\end{eqnarray}
\end{widetext}

Taking the zero $T$ limit, we obtain

\begin{eqnarray}
\label{AntiSymGE1}
\frac{\Delta_{a}}{g_a} &=& \frac{1}{8} \int \frac{d k~k^2}{2 \pi^2}   \big[ \frac{\Delta_{a}}{E_{\textbf{k}, 1}} + \frac{\Delta_{a}}{E_{\textbf{k},2} } + \left( \frac{1}{E_{\textbf{k}, 1}}\!- \frac{1}{E_{\textbf{k}, 2}}\right)\\
\nonumber
&\times&  \frac{\Delta_{a} (\Delta_{a}^2 - \Delta_{b}^2+ \varepsilon_{a}({\bf k})^{2} - \varepsilon_{b}({\bf k})^{2}) + 2 |V_{{\bf{k}}}|^{2}(\Delta_{a} + \Delta_{b})}{E_{\textbf{k}, 1}^{2} - E_{\textbf{k}, 2}^{2}}  \big],\\
\frac{\Delta_{b}}{g_b} &=& \frac{1}{8} \int \frac{d k~k^2}{2 \pi^2}  \big[ \frac{\Delta_{b}}{E_{\textbf{k}, 1}} + \frac{\Delta_b}{E_{\textbf{k},2} } - \left( \frac{1}{E_{\textbf{k}, 1}}\!- \frac{1}{E_{\textbf{k}, 2}}\right)\\
\nonumber
&\times&  \frac{\Delta_{b} (\Delta_{a}^2 - \Delta_{b}^2+ \varepsilon_{a}({\bf k})^{2} - \varepsilon_{b}({\bf k})^{2}) + 2 |V_{{\bf{k}}}|^{2}(\Delta_{a} + \Delta_{b})}{E_{\textbf{k}, 1}^{2} - E_{\textbf{k}, 2}^{2}}  \big].
\end{eqnarray}

\end{document}